\documentclass[pre,aps,twocolumn,showpacs]{revtex4}
\usepackage{graphicx}
\usepackage{amsmath,bm}
\begin{document}
\pagestyle{plain}

\title{A Minimal Stochastic Model for Fermi's Acceleration}

\author{Freddy Bouchet}
\affiliation{Dipartimento di Fisica Universit\`a "La Sapienza" 
P.le A.~Moro 2, I-00185 Roma, Italy}
\author{Fabio Cecconi}
\affiliation{Dipartimento di Fisica Universit\`a "La Sapienza" \& INFM UdR 
Roma-1 and Center for Statistical Mechanics and Complexity (S.M.C.), 
P.le A.~Moro 2, I-00185 Roma, Italy}
\author{Angelo Vulpiani}
\affiliation{Dipartimento di Fisica Universit\`a "La Sapienza" \& INFM UdR
Roma-1 and S.M.C., P.le A.~Moro 2, I-00185 Roma, Italy}
\begin{abstract}
We introduce a simple stochastic system able to generate anomalous 
diffusion both for position and velocity. The model represents
a viable description of the Fermi's acceleration mechanism and it is 
amenable to analytical treatment through a linear Boltzmann equation.
The asymptotic probability distribution functions (PDF) for velocity 
and position
are explicitly derived. The diffusion process is highly 
non-Gaussian and the time growth of moments is characterized by only 
two exponents $\nu_x$ and $\nu_v$. The diffusion process is anomalous 
(non Gaussian) but with a defined scaling properties i.e. 
$P(|{\bf  r}|,t) = 1/t^{\nu_x}F_x(|{\bf r}|/t^{\nu_x})$ and similarly 
for velocity.
\end{abstract}


\maketitle
\date{\today}
About half a century ago, Fermi proposed an acceleration mechanism for 
interstellar particles, now referred as Fermi's 
acceleration \cite{Fermi}, to explain the very high energy of  
cosmic rays.  Nowadays, Fermi's acceleration remains one of the relevant 
explanations for several phenomena in plasma physics~\cite{Plasma} and  
astrophysics~\cite{Astro1,Astro2}. 
In Fermi's mechanism, at variance with diffusion in real space   
({\em e.g.} Lorentz's gas \cite{Lorentz,Dorfman}), it is the 
velocity that undergoes diffusion due to the presence of a stochastic 
acceleration.  
This original idea~\cite{Fermi} found also a number of applications   
in the theory of dynamical systems, because it offers a simple  
but not trivial way to generate chaotic systems  
\cite{Chirikov,LL,Chaos,Chaos1,Chaos2,Chaos3,Livi} to describe the dynamics of  
comets~\cite{Comets} and the motion of charged particles into  
electromagnetic fields~\cite{Manolakou}. 

This letter studies the Fermi's mechanism through a simple  
model, in which a particle may absorb kinetic energy (accelerate)  
through collisions against moving scatterers. The purpose of the model
is to provide a description of acceleration mechanism without specific 
assumptions on the interstellar media (e.g. presence of turbulent magnetic 
fiels or matter fractal distribution). 
Since we are interested mainly in the general features of the diffusion 
process in phase space, we consider only non-relativistic classical 
particles, neglecting the details of their interactions with magnetic fields.
Our model is able to produce an acceleration process characterized by  
anomalous diffusion both in velocity and position: 
$\langle |{\bf v}(t)|^2 \rangle \sim t^{2\nu_v}$,  
$\langle |{\bf x}(t)|^2 \rangle \sim t^{2\nu_x}$ ($\nu_v>1/2$, $\nu_x>1/2$) 
and non Gaussian distributions. 
For simplicity we restrict to a two dimensional system  
(${\bf x}(t)\in {\bf R}^2$) and circular obstacles with radius $a$.  
These obstacles represent the regions where the magnetic field irregularities,
responsible for the scattering, are localized. We shall assume that they 
are randomly distributed in the plane with uniform density $\rho$. 
Their velocity ${\bf V} = (V\cos\phi,V\sin\phi)$ is chosen according 
to an isotropic probability $G(V)dVd\phi/2\pi$, with the only 
requirement that $\langle V^{2}\rangle < \infty$.  
We furthermore assume, that the influence of the particles on the magnetic 
scatterers is so negligible, that the velocity of the latter 
is kept unchanged during collisions. 
In the reference frame of the scatterers, where the energy of the particle  
remains constant, the collision is completely determined by the trajectory  
deflection angle and collision time. Such a deflection angle 
$\alpha = \alpha({\bf V},{\bf v},b,H)$ depends on particle 
and obstacle velocity, on the impact parameter $b$ and on the 
shape of the magnetic field $H$ inside the obstacle.  
 
After a collision, the new particle velocity is  
${\bf v}_{S}^{A} = {\bf M}(\alpha) {\bf v}_{S}^{B}$, 
where the superscripts $A$ and $B$ stand for ``after'' and ``before'' 
the collision respectively, while subscript $S$ refers to  
velocities in the scatterers reference frame and the rotation  
matrix ${\bf M}(\alpha)$  
describes the scattering deflection.  
In the fixed reference frame, ${\bf v}={\bf v}_S+{\bf V}$, we have: 
$$ 
{\bf v}^{A}={\bf V} + {\bf M}(\alpha)({\bf v}^{B}-{\bf V})\,. 
$$ 
A further simplification occurs when the density of the  
scatterers $\rho$ 
is small. Indeed, in the low density limit 
($\rho \sim l_{0}^{-2} \ll a^{-2}$),  
the mean free path is much larger than the average distance between 
nearest obstacles $l_0$ and two successive collisions can be safely  
assumed to be independent. 
Such a simplification allows describing the evolution 
of the system as a stochastic map: 
\begin{equation} 
\begin{array}{rcl} 
{\bf r}_{n+1} & = & {\bf r}_{n}+{\bf v}_{n}\Delta t_{n}, \quad\quad 
t_{n+1}  = t_{n}+\Delta t_{n}\\ 
{\bf v}_{n+1} & = & {\bf V}+ {\bf M}(\alpha)({\bf v}_{n}-{\bf V})  
\end{array} 
\label{eq:Map} 
\end{equation} 
where $\Delta t_{n}= \ell_{n}/|{\bf v}_{n}|$ and $\ell_{n}$ are 
the time and the free path between two consecutive collisions, 
respectively. 
Under the hypothesis of independent collisions,  
well verified at low densities, $\alpha$ and $\Delta t_{n}$ become   
independent random variables, whose distribution will be determined by  
the distribution of the obstacle velocities ${\bf V}$ and impact 
parameters $b$. 
For sake of simplicity, we shall suppose that the 
probability distribution $Q(\alpha)$ for $\alpha$  
is independent of ${\bf V}$ and is an even function (for symmetry  
reasons).  
 
Let us now specify, the probability distribution of $\Delta t$. 
If the scatterers were at rest, a   
particle with speed $v$ would encounter uniformly distributed  
obstacles, then the variable $\Delta t$  
would follow an  
exponential distribution law: 
$P(\Delta t) = \lambda \exp (-\lambda \Delta t)$ 
with a decay rate $\lambda = 2 a\rho {\bf v}$. 
However, in the case of moving obstacles, a particle will  
encounter preferentially those obstacles with a velocity direction 
opposite to its own velocity. 
Then, the rate $\lambda$ has to be replaced by  
$\lambda({\bf {v}},{\bf V})= 2a \rho |{\bf v}-{\bf V}|$.  
Therefore the probability that a particle 
with velocity ${\bf v}$ undergoes a collision with an obstacle 
of velocity ${\bf V}$, after a time $\Delta t$  
from the previous collision is:  
\begin{equation} 
P_{{\bf V}}(\Delta t|{\bf v})= 2a \rho |{\bf v}-{\bf V}|G(V) 
\exp \{-\bar{\lambda}({\bf v})\Delta t\}\,, 
\label{eq:Probcoll} 
\end{equation} 
with,  
$ 
\bar{\lambda}({\bf v})=2a\rho \langle |{\bf v}-{\bf V}|\rangle_{{\bf V}}$  
where $\langle ....\rangle_{{\bf V}}$ indicates  
the average on ${\bf V}$ \cite{Note_Proba}. 
Of course more general laws can be considered  
to take into account the presence of a minimal collision  
time, but our simulations show that  
the results are fairly independent of the details of $P(\Delta t)$.

The above system might look somehow trivial when considering its 
statistical properties as function of the number of collisions.  
Noting that ${\bf v}_{n+1}$ is obtained from   
${\bf v}_{n}$ by a random rotation plus a random shift,
one can expect, from the central limit theorem, that  
$\langle |{\bf v}_{n}|^{2} \rangle \sim n$ and analogously 
$\langle |{\bf r}_{n}|^{2} \rangle \sim n$, of course 
the PDF of ${\bf v}_{n}$ and ${\bf }_{n}$ are Gaussian.  
However, the proper question is about the dependence in time rather
than in the $n$, so the particles at the same time $t$
experience a different number of collisions and therefore  
the PDF of ${\bf v}(t)$ and ${\bf r}(t)$ are not expected to be Gaussian.  
The time after $n$ collisions,
$t = \sum_{k=1}^n \Delta t_{k}$ is the sum of random quantities
and its dependence on $n$ can be worked out via the following scaling 
argument~\cite{Redner}.
Since  
$\Delta t_k \sim \ell / \sqrt{\langle |{\bf v}_k|^2 \rangle} \sim 1/\sqrt{k}$,
then $t \sim\sum_{k=1}^n k^{-1/2} \sim \sqrt{n}$;
$\langle |{\bf r}_{n}|^{2} \rangle \sim n$ correspond to the
scaling  laws: 
\begin{equation} 
\langle|{\bf v}(t)|^2 \rangle \sim t^{2}  
\quad\quad \mbox{and} \quad\quad   
\langle|{\bf r}(t)|^2 \rangle \sim t^2\;. 
\label{eq:ballistic} 
\end{equation} 
In the following, we will show that this ``mean field'' scaling behaviour 
is in fact correct.  
 
Let us now present a simple but relevant remark: an easy computation 
gives  
\begin{equation} 
\label{Mean_Square_displacement} 
\langle |x(t)-x(0)|^2 \rangle \simeq 2\int _{0}^{t}\int _{0}^{t-t'} 
\langle |v_x(t')|^{2} \rangle R_x(t',\tau ) dt'd\tau  
\end{equation} 
with 
$ 
R_x(t,\tau )=\langle v_x(t) v_x(t+\tau )/   
\langle |v_x(t)|^{2} \rangle  
$. 
In Lorentz-gas with fixed obstacles and elastic collisions, 
$\langle |v_x(t')|^{2} \rangle$ remains bounded, thus the integration over 
$t'$ yields a proportionality to $t$.  
In addition stationarity makes $R_{i}(t',\tau)$ independent of $t'$    
and an anomalous diffusion can originate only by long time 
tail of $R_{i}(\tau)$ (i.e. slower than $O(\tau ^{-1})$).  
 
By contrast, in systems with Fermi's acceleration which are non 
stationary, $R_x(t,\tau)$ depends also on $t$, and the scaling laws  
$\langle|{\bf v}(t)|^2\rangle \sim t^{2\nu_v}$ and  
$\langle|{\bf r}(t)|^2\rangle \sim t^{2\nu_x}$, 
could not be trivially related (see~\ref{Mean_Square_displacement}). 
In this case, one can only derive the bound:  
$$ 
\nu_{x} \leq \nu_{v}+1\;, 
$$ 
the equality holding in presence of a very strong time 
correlation among the velocities.  
 
A Boltzmann-like description of our model can be carried out  
under the hypothesis of the independence of consecutive collisions,  
already employed to write the dynamics (\ref{eq:Map},\ref{eq:Probcoll}).  
The spatial homogeneous probability $f({\bf v},t)$, that a particle  
has a velocity ${\bf v}$ at time $t$, evolves according to the equation 
\begin{equation} 
\label{Boltzman} 
\partial_{t}f({\bf v},t) = -\bar{\lambda}({\bf v}) f({\bf v},t) +  
\int d{\bf u}\, T({\bf v},{\bf u}) f({\bf u},t) \, , 
\end{equation} 
where, from Eqs.~(\ref{eq:Map}) and (\ref{eq:Probcoll}),  
the transition probability $T$ reads  
\begin{equation} 
\label{eq:T} 
T({\bf v},{\bf u}) = \langle \lambda({\bf u},{\bf V})  
\delta[{\bf v} - {\bf V} - {\bf M}(\alpha)(  
{\bf u} - {\bf V})]  \rangle_{{\bf V},\alpha}\;. 
\end{equation} 
Equation~(\ref{Boltzman}), at variance with the usual Boltzmann equation, 
is linear because of the independence between obstacles and particles. 
Moreover, the isotropy of the scatterer distribution insures that $T$ is 
only a  function of the moduli $v$,$v^B$ and of the angle $\theta-\theta^B$ 
between ${\bf v}$ and ${\bf v}^{B}$: 
$T=T(v,v^{B},\theta - \theta^B)$. Therefore it is convenient to expand $f$ in  
Fourier series of the angular variable: 
$f(v,\theta) = \sum_{k}f_{k}(v) \exp(ik\theta)$. 
The linearity of eq.~(\ref{Boltzman}) decouples Fourier modes 
$f_{k}$ and we obtain  
$\partial _{t}f_{k}(v,t) =-\lambda(v)  
f_{k}+ \int dv^{B}\, T_{k}(v,v^{B})  
f_{k}(v^{B},t)$ 
with $T_{k}(v,v^{B}) = 
\int d\theta \, \exp(ik\theta) T(v,v^{B},\theta)$. 
 
We can restrict to the asymptotic time behavior because the particles 
accelerate and their distribution is rapidly dominated by  
velocities $v\gg V$. 
Then an asymptotic expansion of the operator $T$ is possible in the small 
parameter $V/v$. 
 
We first consider the evolution of an isotropic density $f(v,\theta)=f(v)$ 
by introducing the PDF of the velocity modulus, $v=|{\bf v}|$,  
$g(v,t) = 2\pi v f(|{\bf v}|)$. We can then substitute    
$\delta[{\bf v} - {\bf V} - {\bf M}(\alpha)({\bf v}^{B} - {\bf V})]$ with  
$\delta[{\bf v}^{B} - {\bf v} - (1-M(\alpha)) {\bf V}]$  in  
Eq.~(\ref{eq:T}) and, at second order in $v/V$, 
we obtain~\cite{Calcul}: 
\begin{equation} 
\label{Fokker_Planck_g} 
\partial_{t} g(v,t) = D v \partial^2_v g\;, 
\end{equation} 
where $D = a\rho \sigma_{F}\langle V^{2}\rangle_{\bf V}$ and 
$\sigma_{F} = \langle 1-\cos\alpha \rangle_{\alpha}$. 
The asymptotic solution of this Fokker-Planck equation may be explicitly  
derived~\cite{Calcul} and it converges toward the scaling  
function $g_{a}(v,t) = h[v/(Dt)]/(Dt)$. A direct substitution into 
Eq.~(\ref{Fokker_Planck_g}) shows  
that the scaling function $h$ verifies the equation:  
$\xi h^{''}(\xi) +\xi h^{'}(\xi) + h(\xi) =0$, whose solution is  
$h(\xi) = \xi\exp(-\xi)$.   
Thus $g_{a}(v,t) = v/(Dt)^2\exp\{-v/(Dt)\}$ and the moments of $v$  
are given by $\langle v^{n} \rangle \sim (n+1)! (Dt)^{n}$. 
 
We now consider the evolution of a non isotropic Fourier mode  
$f(v,\theta) = \exp(ik\theta) f_{k}(v)$.  
We can then substitute  
$\delta[{\bf v} - {\bf M}(\alpha){\bf v}^{B} - (1 - {\bf M}(\alpha)) {\bf V}]$  
with  
$\exp(-ik\alpha)\delta[{\bf v}^{B}-{\bf v}-(1-M(\alpha)) {\bf V}]$,  
in Eq.~(\ref{eq:T}) and performing the expansion at small $v/V$  
we obtain  
\begin{equation} 
\label{Relaxation_anisotropic}  
\partial_{t}f_{k}(v) = -L_k f_{k}(v)  
\end{equation} 
at leading order in $V/v$, with $L_k = 2a\rho \sigma_{F,k}$ and 
$\sigma_{F,k}=  \langle 1-\cos(k\alpha) \rangle_{\alpha}$ 
(we have exploited even symmetry of the $\alpha$ distribution $Q$).  
Thus for any initial distribution function, $f_{k}$ relaxes exponentially.  
This is physically clear, as few collisions randomize the phase of the 
velocity. The proportionality to $v$ of the relaxation rate  
is a consequence of the fact that the number of collisions per unit 
time is proportional to $v$.  
We note that the neglected higher order terms in $V/v$, in  
Eq.~(\ref{Relaxation_anisotropic}), are a transport and a diffusion term,  
not relevant for large $v$. 
 
This result shows that any distribution rapidly becomes isotropic, it  
also allows to compute the velocity autocorrelation function and some 
properties of the spatial diffusion.  
For large $t$, the velocity autocorrelation can be expressed in terms of  
the asymptotic distribution $g_a$ and of the solution of the Boltzmann  
equation (\ref{Boltzman}), with a delta-like initial condition 
$\delta({\bf v}'-{\bf v})$.  
Using Eq.~(\ref{Relaxation_anisotropic}), we can compute this solution.  
Using this result, we obtain the expression for the autocorrelation function:  
\begin{equation} 
\label{autocorrelation} 
\langle v_{x}(t) v_{x}(t+\tau)\rangle = 3(Dt)^{2}/(1 + DL_1t\tau)^{4}. 
\end{equation}
The diffusion properties of velocity can be derived directly by  
Eq.~(\ref{autocorrelation}) computed at $t=0$ and from the simmetry
$x\leftrightarrow y$
\begin{equation} 
\label{diffusion} 
\langle |{\bf v}(t)|^2 \rangle = 6(Dt)^2  
\end{equation} 
The algebraic decay of the function~(\ref{autocorrelation}) is fast enough 
to make the integral~(\ref{Mean_Square_displacement}) convergent then    
\begin{equation} 
\label{diffux} 
\langle |x(t)-x(0)|^2 \rangle  
\sim  \frac{2D}{L_1}t^2 
\end{equation} 
The above results~(\ref{autocorrelation},\ref{diffusion},\ref{diffux}) agree with the 
simple argument leading to Eq.~(\ref{eq:ballistic}).  
In order to study the statistical properties of the particle position,  
we consider the evolution of the PDF, $f({\bf r},{\bf v},t)$,   
for the velocity and position.  
The generalization of Eq.~(\ref{Boltzman}) is the inhomogeneous  
Boltzmann equation where $\partial_t f$ is replaced 
by $\partial_{t}f({\bf r},{\bf v},t) +{\bf v}.\nabla_{{\bf r}}f$.   
We introduce the distribution  
$g(r,\phi,v,\theta,t) = 4\pi^{2}rv f({\bf r},{\bf v},t)$ in polar  
coordinates, ${\bf r}=(r\cos \phi,r\cos \phi)$   
${\bf v}=(v\cos \theta,v\cos \theta)$.  
We limit the discussion to the isotropic case, that is  
$g(r,\phi,v,\theta) = g(r,v,\theta -\phi)$. 
Again, it is convenient to develop the angular dependence in Fourier  
modes: $g(r,\phi ,v,\theta) =\sum _{k}g_{k}( r,v)\exp(ik(\theta -\phi ))$. 
 
The evolution equation for each mode $g_k$ can be written down and 
studied in the long time limit, but we do not report here the detailed 
derivation. 
We can prove that $g_0$ is dominant for large $t$ and converges toward a 
scaling function $h_{0}[L_1^{1/2}r/(D^{1/2}t),v/(Dt)]  
L_1^{1/2}/(D^{3/2}t^{2})$ \cite{Cons} which verifies the equation 
\begin{equation} 
\label{eq:Scaling_Position_Velocity} 
2h_{0} + \eta \partial_{\eta}h_{0} + \xi \partial_{\xi}h_{0} +  
\xi\partial^{2}_{\xi}h_{0} + \frac{\xi}{4} 
\left[\partial^{2}_{\eta}h_{0}-\partial_{\eta} 
\left(\frac{h_{0}}{\eta }\right) \right] = 0 
\end{equation} 

We were not able to explicitly solve Eq.~(\ref{eq:Scaling_Position_Velocity}), 
however we computed all the moments of $h_0$ by recurrence. 
Indeed, if we define  
$a_{k,n}=\int^{\infty }_{0}drdv\, v^{k}r^{n}h_{0}(r,v)$, 
we obtain:  
$(n+k)a_{k,n}=k(k+1)a_{k-1,n}+n^{2}/4a_{k+1,n-2}$. 
Using that $\int^{\infty}_{0}d\eta \, h_{0}(\eta,\xi) = h(\xi) =  
\xi \exp (-\xi)$, we have $a_{k,0}=(k+1)!$. The recurrence relation then,  
for instance, gives:  
$\langle r^{2}\rangle = 2\langle|x(t)-x(0)|^{2} \rangle  
\sim ~ Dt^{2}/L_1$, 
 $\langle r^{4}\rangle \sim 8D^{2}t^{4}/(3L_1^{2})$ 
and so on. 
 
\begin{figure} 
\includegraphics[clip=true,width=\columnwidth,keepaspectratio] 
{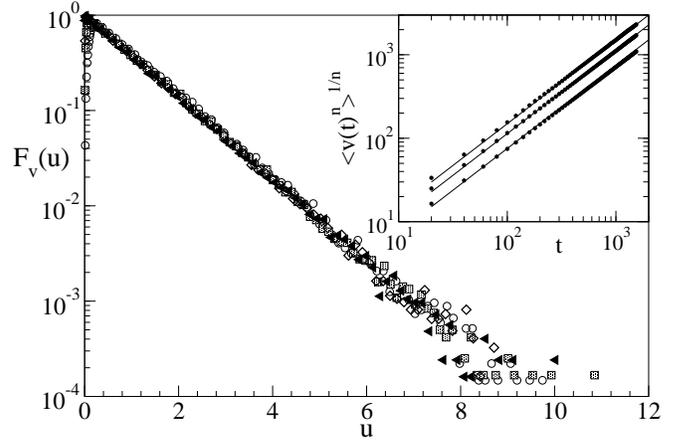} 
\caption{\label{fig:pdfv} PDF of rescaled particle  
velocity $u = |{\bf v}|/\sqrt{\langle v(t)^2 \rangle}$ at times  
$t=80,320,1280,5120$, corresponding to circles, squares, 
diamonds and triangles respectively.  
The perfect collapse is in accordance with the 
theoretical results. 
Inset: $\langle v(t)^n \rangle^{1/n}$ vs $t$, for 
$n=2,4,6$ from bottom to top; straight lines have slope $1$}.
\end{figure}

\begin{figure} 
\includegraphics[clip=true,width=\columnwidth,keepaspectratio] 
{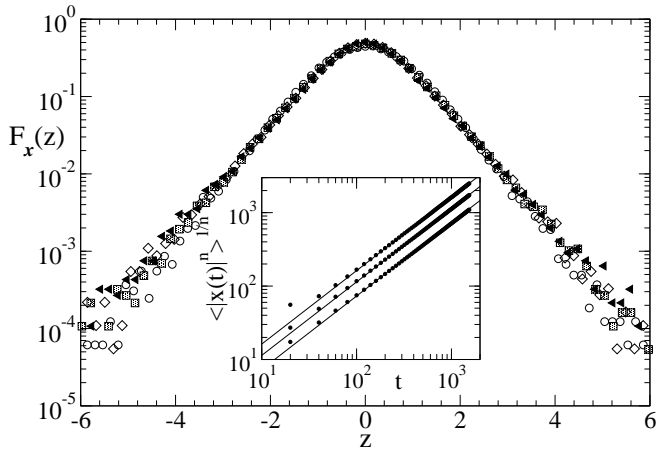} 
\caption{\label{fig:pdfx} PDF of the rescaled coordinate  
$z = x/\sqrt{\langle x(t)^2 \rangle}$ at times   
$t=80,320,1280,5120$ with the same symbols of fig.~\ref{fig:pdfv}. 
Inset: $\langle x(t)^n \rangle^{1/n}$ vs $t$, for 
$n=2,4,6$ from bottom to top; straight lines have slope $1$.}
\end{figure}

Numerical simulations confirm our theoretical results obtained via the  
Boltzmann equation for the scaling of the moments of position and velocity.
Simulations were run for a set of $N=10^6$ particles starting from 
random initial positions and velocities well localized in a region 
of the phase space with size small compared with $l_0^2$ and $V_0^2$. 
For a particle with velocity 
${\bf v}$, we randomly select a scatter velocity ${\bf V}$ from 
$G(V) = \delta(V-V_0)/2\pi$ and a scattering angle $\alpha$ uniformely 
distributed in $[0, 2\pi]$. Then 
the collision time $\Delta t_n$ was drawn from the 
law~(\ref{eq:Probcoll}) and finally particle position and velocity 
were updated via the rule~(\ref{eq:Map}). 
Figs.~\ref{fig:pdfv}, and \ref{fig:pdfx} show, for different times,  
the rescaled PDF for the velocity modulus 
$v=|{\bf v}|$ and $x$-coordinate.  
The collapse of the curves is very good and the exponential tails agree 
with our theoretical predictions. 
The qualitative scenario corresponding to $\nu_x = \nu_v = 1$ and 
exponential tails remains unaffected by using generic form of 
$P_{{\bf V}}(\Delta t,{\bf v})$ provided it keeps the exponential 
decay for large $\Delta t$. 
 
We stress that the diffusion process is anomalous, 
since $\nu_x$ is different from $1/2$ (we discuss only the diffusion for  
the position but similar considerations hold for velocity). 
The existence of a unique exponent characterizing the  
growth of the different moments  
$\langle |{\bf r}(t)|^n \rangle \sim t^{n \nu_x}$ is somehow peculiar, and in  
ref.~\cite{Casti} is called {\em weak anomalous  
diffusion}, because the most general anomalous diffusion   
(strong anomalous diffusion)  implies that  
$\langle|{\bf r}(t)|^n \rangle \sim t^{n \nu_x(n)}$ with $\nu_x(n)$ a non  
constant parameter, therefore all the PDF's cannot be rescaled onto a  
single curve. 

The model can be made more realistic taking into  
account possible energy losses due to interactions with the  
medium and energy irradiation.  This dissipation is mimicked by rescaling  
the velocity components by a factor  
$\sqrt{1 - \gamma}$ after each collision,  
where  $ \gamma = |{\bf v}^A - {\bf v}^B|/[\tau_c (v^A + v^B)] $ 
with $\tau_c$ a typical time of the collision.  
The presence of dissipation introduces neither      
substantial modification on tailed structure of position and velocity PDFs,
nor change their scaling behavior. Even different forms of $\gamma$ do not 
affect the global scenario. 
 
In summary, we introduced a simplified treatable model of Fermi's  
acceleration. It is remarkable that the system presents an  
anomalous (super) diffusion both for position and velocity, which is
robust under changes of the details and it is
characterized by only two exponents $\nu_x=\nu_v=1$ and by   
the PDFs' scaling behavior  
$P(|{\bf r}|, t) = 1/t^{\nu_x} 
F_x(|{\bf r}|/t^{\nu_x}), \, P(|{\bf v}|, t) = 1/t^{\nu_v} 
F_v(|{\bf v}|/t^{\nu_v})$. 
Furthermore, as a consequence of non trivial correlations, 
the two exponents $\nu_x$ and $\nu_v$ are not related by a simple 
dimensional argument.

A.V. and F.C. acknowledge the financial support by Cofin Miur 2001 
{\em Fisica Statistica di Sistemi Classici e Quantistici}.
F.B. is supported by E.U. Network {\em Stirring and Mixing} 
(RTN2-2001-00285).  

 
 

\begin{thebibliography}{99}
\bibitem{Note_Proba}  
Consider the case with two kinds of obstacles with respective 
dacay rates $\lambda_1$ and $\lambda_2$. Let $P_1(t)$ and $P_2(t)$ the 
probability to have an encounter between times $t$ and $t+\Delta t$ for 
events of types $1$ or $2$ respectively. Let $N_i(t) = 1-\int_0^t 
P_i(u)du$ for $i=1,2$, the probability not to have had an encounter of 
type $i$ before $t$. Then $dN_i/dt = -\lambda_i (N_1+N_2)$. Thus 
$d(N_1+N_2)/dt = -(\lambda_1+\lambda_2)(N_1+N_2)$. $N_1+N_2$ is 
exponential with decay rate $\lambda_1+\lambda_2$ and $P_i(t)=\lambda_i 
exp(-(\lambda_1 + \lambda_2)t)$. The generalization of this result to an 
infinite number of rates gives the result (\ref{eq:Probcoll}).
\bibitem{Cons} Using the scaling solution, we have $g_{k}=h_{k}  
(\eta,\xi) /t^{k}$, $\partial_{t}g_{k}=A(\eta ,\xi) 
/t^{k+1}$, 
$L_{k}vg_{k}=A(\eta ,\xi)/t^{k-1}$. Thus the neglected 
terms are of order $1/t^{2}$ times the conserved ones, this is  
a good check of our approximations.
\end{thebibliography}
\end{document}